\documentclass[10pt,aps,showpacs,nofootinbib,prd,aps,epsf,floats,
               amsmath,amssymb,amsfonts,axodraw,floatfix,graphicx,twocolumn]{revtex4-1}

\usepackage{amsmath, amssymb}
\usepackage{axodraw}
\usepackage{multirow}
\usepackage{paralist}
\newcommand{\mathsym}[1]{{}}

\usepackage{graphicx}
\usepackage{amsmath}
\usepackage{amssymb}
\usepackage{amsmath}
\usepackage{multirow}
\usepackage{paralist}
\usepackage{slashed}
\usepackage{amsfonts}
\usepackage{hyperref} 
\usepackage{graphicx}
\usepackage{amsmath}
\usepackage{amssymb}
\usepackage{mathrsfs}
\newsavebox{\PSLASH}
 \sbox{\PSLASH}{$p$\hspace{-1.8mm}/}
 
\renewcommand{\theequation}{\thesection.\arabic{equation}}
\newcounter{saveeqn}
\newcommand{\add}{\addtocounter{equation}{1}}
\newcommand{\alphaeqn}{\setcounter{saveeqn}{\value{equation}}%
\setcounter{equation}{0}%
\renewcommand{\theequation}{\mbox{\thesection.\arabic{saveeqn}{\alpha{equation}}}}}
\newcommand{\reseteqn}{\setcounter{equation}{\value{saveeqn}}%
\renewcommand{\theequation}{\thesection.\arabic{equation}}}

 \newsavebox{\notrightarrow}
 \sbox{\notrightarrow}{$\to$\hspace{-4mm}/}
 
 \newsavebox{\PARTIALSLASH}
 \sbox{\PARTIALSLASH}{$\partial$\hspace{-1.6mm}/}
 
 \newsavebox{\ASLASH}
 \sbox{\ASLASH}{$A$\hspace{-2.1mm}/}
 
 \newsavebox{\KSLASH}
 \sbox{\KSLASH}{$k$\hspace{-1.8mm}/}
 
 \newsavebox{\LSLASH}
 \sbox{\LSLASH}{$\ell$\hspace{-1.8mm}/}
 
 \newsavebox{\QSLASH}
 \sbox{\QSLASH}{$q$\hspace{-1.8mm}/}
 
 \newsavebox{\DSLASH}
 \sbox{\DSLASH}{$D$\hspace{-2.2mm}/}
 
 \newsavebox{\DbfSLASH}
 \sbox{\DbfSLASH}{${\mathbf D}$\hspace{-2.8mm}/}
 
 \newsavebox{\DELVECRIGHT}
 \sbox{\DELVECRIGHT}{$\stackrel{\rightarrow}{\partial}$}
 
 \newcommand{\blue}{\IfColor{\textCadetBlue}{}}
\newcommand{\black}{\IfColor{\textBlack}{}}
\newcommand{\red}{\IfColor{\textRed}{}}
\newcommand{\green}{\IfColor{\textOliveGreen}{}}
\newcommand{\lil}{\IfColor{\textRedViolet}{}}

\newcommand{\bs}{\boldsymbol}

\makeatother
\usepackage[T1]{fontenc}
\usepackage[latin9]{inputenc}
\usepackage{amsmath}
\usepackage{amssymb}
\usepackage{graphicx}
\usepackage{dcolumn}
\usepackage{verbatim}
\usepackage{orcidlink}

\begin{document}
\title{Relativistic Barnett effect and Curie law in a rigidly rotating free Fermi gas}
\author{M. Abedlou Ahadi\ \ }\email{mohammad\_abedlouahadi@physics.sharif.ir}\author{N. Sadooghi\,~\orcidlink{0000-0001-5031-9675}~~}\email{CA: sadooghi@physics.sharif.ir, neda.sadooghi@gmail.com}
\affiliation{Department of Physics, Sharif University of Technology,
P.O. Box 11155-9161, Tehran, Iran}
\begin{abstract}
By combining methods from thermal field theory and statistical mechanics, we reexamine the spin polarization caused by the relativistic Barnett effect in a rigidly rotating Fermi gas. We determine the pressure of this medium and show that it depends on an effective chemical potential, which includes contributions from orbital angular momentum-rotation and spin-rotation coupling. We introduce a specific regularization scheme to sum over the angular momentum quantum numbers. As a result, the thermal pressure and all thermodynamic quantities are separated into two parts that differ only in the spin fugacities of spin-up and spin-down fermions. We calculate the Fermi energy for both components and show that the Fermi energy of the spin-down fermions is lower than that of the spin-up ones. This difference arises from the spin-rotation coupling and leads to a spin polarization consistent with the Barnett effect. In particular, we introduce the spin-chemicorotational ratio $\eta\equiv \Omega^{(0)}/2\mu^{(0)}$, which adjusts the spin polarization of the Fermi gas. Here, $\Omega^{(0)}$ and $\mu^{(0)}$ represent the angular velocity and chemical potential at zero temperature, respectively. The factor $1/2$ accounts for the fermion's spin. We explore the temperature dependence of $\mu$ and $\Omega$, while assuming that the number of spin-up and spin-down fermions remains temperature independent. Our findings indicate that the spin-down component of the rotating Fermi gas dilutes at lower temperatures compared to the spin-up component. Additionally, we calculate the magnetic susceptibility arising from the Barnett magnetization and demonstrate that it is proportional to the moment of inertia $I$ of the rotating Fermi gas. Finally, we prove that $I$ exhibits a $1/T$ behavior in the high-temperature limit, similar to the Curie law of paramagnetism.
\end{abstract}
\maketitle
\section{Introduction}\label{sec1}
\setcounter{equation}{0}
One intriguing question in many-body systems of fermions and bosons is how external electromagnetic fields and rotation affect their thermodynamic properties. The implications of these phenomena extend across various branches of physics, from nonrelativistic condensed matter physics to ultrarelativistic heavy ion collisions. Intensive experiments are currently in progress at the Relativistic Heavy Ion Collider (RHIC) and the Large Hadron Collider (LHC) to better understand the nature of the matter produced after ultrarelativistic heavy ion collisions (HICs) \cite{rajagopal2018, pisarski2022, aarts2023, hot-QCD, present2023, becattini2021}. In noncentral HICs, large magnetic fields, ranging from $10^{18}$-$10^{20}$ Gau\ss, are generated by electric currents produced from the accelerated motion of positively charged spectator nucleons that do not participate in the collision \cite{gursoy2018, huang-review, warringa2007, skokov2009}. Additionally, large angular momentum of the colliding nuclei results in extremely high global angular velocities, reaching up to $10^{21}$ rad/s \cite{becattini2016}. These extreme conditions significantly influence the early-time dynamics of the quark-gluon plasma (QGP) formed in these collisions. Beyond the well-known effects of large magnetic fields, such as chiral magnetic effect \cite{fukushima2008, kharzeev2015} and magnetic catalysis as well as inverse magnetic catalysis, that modify the phase diagram of quantum chromodynamics (QCD) \cite{fayazbakhsh2010, fayazbakhsh2011, rebhan2011, bali2012-1, bali2012-2, delia2013, bruckmann2013, fayazbakhsh2014, ayala2014,shovkovy2015,cao2021}, extreme rotation similarly affects both the thermodynamic \cite{chernodub2017a, chernodub2017b,zahed2017, fukushima2019, ambrus2019,huang2020, braguta2023a, braguta2023b, ambrus2023,huang2023,siri2024a} and transport properties \cite{kharzeev2015} of the QCD matter,  as well as its phase structure \cite{yamamoto2013,mameda2015,fukushima2015, braguta2021, chernodub2021,sadooghi2021,sadooghi2023,cao2023,sun2024, siri2024b, singha2024, abedlou2025,singha2025,siri2025,kiamari2025,sahoo2025a,sahoo2025b}.
\par
One of the significant effects associated with extreme rotation is the polarization of $\Lambda$ hyperons, which is observed in HICs \cite{lambda-ex} (for a recent review, see \cite{becattini-book}). Theoretically, the polarization induced by rotation is related to a spin-rotation coupling that aligns spinful particles with the axis of rotation. This effect was discovered by S. Barnett in 1915 \cite{barnett-2,barnett-3,barnett-4,barnett-5}.  Known as the Barnett effect, it refers to the magnetization that occurs due to the mechanical rotation of any object \cite{fukushima2018}. The Barnett effect arises from the conservation of total angular momentum $\bs{J}=\bs{L}+\bs{S}$ and an $\bs{L}\cdot\bs{S}$ coupling. Any changes in the spin $\bs{S}$ must be compensated by an adjustment in the orbital angular momentum $\bs{L}$. Macroscopically, the rotation of any material with a finite magnetic moment $\bs{\mu}$ (which is proportional to spin $\bs{S}$) produces an effective magnetic field $\bs{B}_{\text{eff}}$, that is parallel to the angular velocity $\bs{\Omega}$. On a microscopic level, comparing the potential energies associated with spinful charged particles in a magnetic field $U_{B}\sim \bs{S}\cdot\bs{B}$, and  the energy related to the spin-rotation coupling, $U_{\Omega}\sim \bs{S}\cdot \bs{\Omega}$, leads to $\bs{B}_{\text{eff}}\sim \bs{\Omega}$. The magnetization resulting from $\bs{B_{\text{eff}}}$ is referred to as the Barnett magnetization $\bs{M}_{\text{Barnett}}$. The Barnett effect, along with its inverse phenomenon, the Einstein de-Haas effect, has many applications in condensed matter physics (for a review of these applications see \cite{fukushima2018} and the papers therein). In \cite{fukushima2018}, an initial attempt is made to present a relativistic generalization of the Barnett effect within the framework of chiral kinetic theory. Recently, the magnetization arising from the nuclear Barnett effect was observed for the first time in a rotating water sample at angular velocities of 13.5 kHz \cite{arabgol2019}. Motivated by this experimental result, the potential consequences of the Barnett effect, which may arise from the enormous $\bs{\Omega}$ created in HICs, are studied in \cite{sahu2026}. Using a rotating Hadron Resonance Gas model, it is demonstrated that the effective magnetic field resulting from the Barnett effect is comparable in strength to the magnetic field produced by accelerated spectator nucleons in HICs. Moreover, the resulting Barnett magnetization is computed and shown to monotonically increase with temperature, baryochemical potential, and angular velocity.
\par
In this paper, we reexamine the Relativistic Barnett effect and its implications. We combine specific concepts from thermal field theory with various aspects of statistical mechanics. This approach results in a novel definition of spin polarization, which is based on distinct features of the model, including the angular velocity and the chemical potential of the rotating medium at zero temperature. The spin-rotation coupling is also essential to this definition, as discussed below. We introduce rotation using a metric originally presented in \cite{yamamoto2013} to describe  rigid rotation. This same metric is employed in recent studies to explore how rotation influences the thermodynamic properties of free bosonic and fermionic systems,  as noted earlier.
\par
We start with the Lagrangian density of Dirac fermions in curved spacetime, described by the metric of rigid rotation with the angular velocity $\bs{\Omega}=\Omega \bs{e}_{z}$. Following the standard imaginary-time formalism of thermal field theory, we determine the pressure of a free Fermi gas under rigid rotation. We show that it depends, as expected, on an effective chemical potential $\mu_{\pm,\ell}\equiv \mu+(\ell\pm 1/2)\Omega$, where $\ell$ is the quantum number corresponding to the third component of the angular momentum, $1/2$ the spin of fermions and $\Omega$ the angular velocity of the rigid rotation. We introduce a specific regularization scheme to sum over $\ell$, and show that the pressure of the system, together with all thermodynamic quantities arising from it, separate into two parts characterized by the fugacity $z_{\pm}\equiv \exp\left(\beta\left(\mu\pm\Omega/2\right)\right)$ corresponding to spin-up ($+$) and spin-down ($-$) fermions. Here, $\beta\equiv 1/T$ is the inverse temperature. The free rotating Fermi gas thus includes
two different components which differ only in their spin fugacity $e^{\pm\beta\Omega/2}$.
Focusing only on the thermal part of their pressure, we determine them in two nonrelativistic (NR) and ultrarelativistic (UR) limits. We show that the spin-up and spin-down component of the Fermi gas behaves differently at zero temperature because their Fermi energies $\epsilon_{F,\pm}$ are different. As it turns out $\epsilon_{F,-}<\epsilon_{F,+}$. We utilize standard methods from statistical mechanics and show that the difference between $\epsilon_{F,-}$ and $\epsilon_{F,+}$ is given by the "spin-chemicorotational ratio" $\eta\equiv \Omega^{(0)}/2\mu^{(0)}$, where $\Omega^{(0)}$ and $\mu^{(0)}$ are the angular velocity and chemical potential at $T=0$. The factor $\eta$ also controls the spin polarization $\mathcal{P}$ and vice versa. The spin polarization is defined by $\mathcal{P}\equiv (n_{+}-n_{-})/(n_{+}+n_{-})$, where $n_{\pm}$ are the number density of spin-up and spin-down fermions.
\par
By assuming that $n_{\pm}$ are temperature-independent, we derive a differential equation for the fugacities $z_{\pm}$. We solve this equation numerically to determine the $T$ dependence of $\mu_{\pm}\equiv \mu\pm \Omega/2$. Our analysis reveals that, depending on the sign of $\mu_{\pm}$, the rotating Fermi gas exhibits three distinct temperature regimes. In the low-temperature regime, both components of the rotating Fermi gas are strongly degenerate. In the intermediate-temperature regime, the spin-up component of the gas remains strongly degenerate while the spin-down component is weakly degenerate. In the high-temperature regime, however, both components become dilute. We show that because of the Barnett effect, the spin-down component of the gas dilutes at a lower temperature than its spin-up component. By applying methods from statistical mechanics, we derive analytical expressions for the $T$ dependence of $\mu_{\pm}$ in these three regimes.
\par
Another intriguing result is related to the $T$ dependence of the moment of inertia of the rotating Fermi gas, particularly at high temperatures. We use the $(T,\Omega)$ dependence of the pressure and determine the $T$ dependence of the angular momentum density $J$ and the moment of inertia $I$. Our results indicate that $I$ decreases as temperature increases, following a $1/T$ behavior in the high-temperature limit. We compare the $T$ dependence of the magnetic susceptibility $\chi_{m}=M_{\text{Barnett}}/B_{\text{eff}}$, which arises from Barnett magnetization with the effective magnetic field $\bs{B}_{\text{eff}}$, with the moment of inertia $I=J/\Omega$, derived from the angular momentum density $J$ and the angular velocity. This comparison reinforces that our conclusion regarding the high-temperature behavior of $I$. In statistical mechanics, this behavior is known as the Curie law of paramagnetism. We present a novel analogy for this phenomenon within the thermodynamic behavior of a rotating Fermi gas.
\par
The organization of the paper is as follows: In Sec. \ref{sec2}, we derive the pressure of a rigidly rotating Fermi gas and, focusing on the expression of the number density $n$ at zero temperature, determine the relation between $\mu^{(0)}$ and $n$ in both NR and UR limits. In Sec. \ref{sec3A}, we discuss the relativistic Barnett effect by introducing the spin polarization $\mathcal{P}$ and show its dependence on spin-chemicorotational ratio $\eta$.
 In Sec. \ref{sec3B}, we determine the $T$ dependence of $\mu$ and $\Omega$ numerically and present analytical expressions for their $T$ dependencies. Finally, in Sec. \ref{sec3C}, we explore the relativistic Curie effect of the moment of inertia of a rigidly rotating Fermi gas. Section \ref{sec4} is devoted to our concluding remarks. In Appendix \ref{appA}, we present the above-mentioned regularization method leading to the summation over the quantum numbers $\ell$ and separate the contributions of spin-up and spin-down fermions to the pressure of the rotating Fermi gas.
\section{Rigidly rotating fermions}\label{sec2}
\setcounter{equation}{0}
We start with the Lagrangian density of free Dirac fermions $\psi$,
\begin{eqnarray}\label{N1}
\mathcal{L} = \bar{\psi} \big[i\gamma^{\mu}\nabla_{\mu} -m +\mu\gamma^{0} \big]\psi,
\end{eqnarray}
where $m$ is the mass and $\mu$ is the chemical potential of the medium. The covariant derivative $\nabla_{\mu}$ is defined by
\begin{eqnarray}\label{N2}
\nabla_{\mu} \psi = (\partial_{\mu} + \Gamma_{\mu} ) \psi.
\end{eqnarray}
Here, the spin connection $\Gamma_{\mu}$ is given by
$\Gamma_{\mu} = -\frac{i}{4} \omega_{\mu i j} \sigma^{ij}
$ with $\omega_{\mu i j}\equiv g_{\alpha \beta} e^{\alpha}_{\,i} (\partial_{\mu} e^{\beta}_{\, j} + \Gamma^{\beta}_{\mu\nu} e^{\nu}_{\, j})$ and $\sigma^{ij} = \frac{i}{2} [\gamma^{i} , \gamma^{j}]$. In $\omega_{\mu i j}$, the tetrads $e_{i}^{\alpha}$, defined by $\eta_{ij}=e_{i}^{\alpha}e_{j}^{\beta}g_{\alpha\beta}$ and the Christoffel symbol, defined by $\Gamma_{\mu\nu}^{\lambda}=\frac{1}{2}g^{\lambda \sigma} (\partial_{\mu}g_{\sigma \nu} + \partial_{\nu}g_{\sigma \mu} -\partial_{\sigma} g_{\mu \nu} )$ are expressed in terms of the metric $g_{\mu\nu}$.
For a rigid rotation around the $z$ axis with the angular velocity $\bs{\Omega}=\Omega\bs{e}_{z}$, $g_{\mu\nu}$ is given by
\begin{equation}\label{N3}
	g_{\mu\nu} =
	\begin{pmatrix}
		1-r^{2}\Omega^{2} & \Omega y & -\Omega x & 0\\
		\Omega y & -1 & 0 & 0 \\
		-\Omega x & 0 & -1 & 0 \\
		0 & 0 & 0 & -1
	\end{pmatrix},
\end{equation}
where $x$ and $y$ are Cartesian coordinates and $r^{2}\equiv x^{2}+y^{2}$. In the above expressions, the Greek and Latin indices $\alpha,\beta\in\{t,x,y,z\}$ and $i,j\in\{0,1,2,3\}$ are the spacetime indices corresponding to the corotating and laboratory frames, respectively. As in  \cite{fukushima2015,chernodub2017a, abedlou2025}, we choose
\begin{eqnarray}\label{N4}
e^{t}_{0} = e^{x}_{1} = e^{y}_{2} = e^{z}_{3} = 1, \quad e^{t}_{1} = y\Omega, \quad  e^{t}_{2} = - x \Omega,\nonumber\\
\end{eqnarray}
and arrive at
\begin{eqnarray}\label{N5}
	\omega_{t12} = - \omega_{t21} = \Omega.
\end{eqnarray}
The only nonvanishing component of the spin connection is thus given by $\Gamma_{t} = - \frac{i}{2} \Omega \sigma^{12}$. Using the metric \eqref{N2}, the nonvanishing components of the Christoffel symbols are given by
\begin{eqnarray}\label{N6}
\Gamma^{x}_{tt} = -x\Omega^{2}  ,&\qquad& \Gamma^{y}_{tt} = -y\Omega^{2}  ,\nonumber\\
\Gamma^{y}_{tx} = \Gamma^{y}_{xt} = \Omega   ,&\qquad& \Gamma^{x}_{ty} = \Gamma^{x}_{yt} = -\Omega.
\end{eqnarray}
Plugging these expressions into \eqref{N1}, the Lagrangian density of rotating fermions thus reads
\begin{eqnarray}\label{N7}
\mathcal{L}=\bar{\psi}[\gamma^{0}\left(i\partial_{t}+\mu+\Omega J_{z}\right)+i\bs{\gamma}\cdot \bs{\nabla}-m]\psi,
\end{eqnarray}
where the $z$ component of the total angular momentum is given by $J_{z}\equiv L_{z}+S_{z}$. Here, $L_{z}\equiv -i\left(x\partial_{y}-y\partial_{x}\right)$ is the angular momentum and $S_{z}\equiv \sigma^{12}/2$, with $\sigma^{12}\equiv \mathbb{I}_{2\times 2}\otimes \sigma^{3}$, where $\mathbb{I}_{2\times 2}=\text{diag}(1,1)$, and $\sigma^{3}$ is the third Pauli matrix. Following the method presented in \cite{abedlou2025}, it is possible to determine the solution of the Dirac equation arising from \eqref{N7}. We arrive after some computation at
\begin{eqnarray}\label{N8}
\psi&=&\left(
\begin{array}{c}
\left(\mathcal{E}^{-}+ik_{\perp}\right)J_{\ell}(u)\\
\left(\mathcal{E}^{+}-ik_{\perp}\right)e^{i\varphi}J_{\ell+1}(u)\\
-\left(\mathcal{F}^{-}+ik_{\perp}\right)J_{\ell}(u)\\
-\left(\mathcal{F}^{+}-ik_{\perp}\right)e^{i\varphi}J_{\ell+1}(u)
\end{array}
\right)e^{-iEt+i\ell\varphi+ik_{z}z},\nonumber\\
\end{eqnarray}
with $\mathcal{E}^{\pm}\equiv \tilde{E}+m\pm k_{z}$, $\mathcal{F}^{\pm}\equiv \tilde{E}-m\pm k_{z}$, $u\equiv k_{\perp}\rho$, and $\tilde{E}\equiv E+\mu+j\Omega$ with $j=\ell+1/2$.
Using this solution, the mode expansion of $\psi(x)$ at finite temperature $T$ within the imaginary-time formalism is given by
\begin{eqnarray}\label{N9}
\psi(x)&=&V^{1/2}\sum_{n,\ell=-\infty}^{+\infty}\int d\tilde{p}~e^{i\left(\omega_{n}\tau+\ell\varphi+p_{z}z\right)}\nonumber\\
&&\times J_{\ell}\left(p_{\perp}r\right)\psi_{n,\ell}(p),
\end{eqnarray}
where
\begin{eqnarray}\label{N10}
\int d\tilde{p}\equiv \int \frac{p_{\perp}dp_{\perp}dp_{z}}{(2\pi)^{2}},
\end{eqnarray}
and $\omega_{n}\equiv (2n+1)\pi T$ with $n\in\mathbb{Z}$ is the fermionic Matsubara frequency. In \eqref{N9},  we use cylindrical coordinate system $x^{\mu}=(t,x,y,z)=\left(t,r\cos\varphi, r\sin\varphi,z\right)$ with $r$ the radial coordinate, $\varphi$ the azimuthal angle, and $z$ the height of the cylinder. Because of the cylindrical symmetry, the expansion includes the Bessel function $J_{\ell}(p_{\perp}r)$ in the radial direction. The latter is labeled by $\ell$, the quantum number corresponding to $L_{z}$. Following the standard method \cite{kapusta-book}, the mode expansion \eqref{N9} is used to determine the partition function $\mathcal{Z}$,
\begin{eqnarray}\label{N11}
\mathcal{Z}=\int \mathcal{D}\psi\mathcal{D}\bar{\psi}\exp\left(\int d^{4}x\mathcal{L}\right),
\end{eqnarray}
with $\mathcal{L}$ given in \eqref{N7}. Using $P_{\text{tot}}=\ln\mathcal{Z}/\beta V$, with $\beta\equiv 1/T$ and the volume $V$, we arrive at the pressure $P_{\text{tot}}$,
\begin{eqnarray}\label{N12}
P_{\text{tot}}=P_{\text{vac}}+P_{\text{mat}}^{\text{particle}}+P_{\text{mat}}^{\text{antiparticle}}.
\end{eqnarray}
Here, the vacuum part
\begin{eqnarray}\label{N13}
P_{\text{vac}}= 2\sum_{\ell=-\infty}^{+\infty}\int d\tilde{p}~\omega_{p},
\end{eqnarray}
and the matter part, including the contributions from particles and antiparticles, reads
\begin{eqnarray}\label{N14}
P_{\text{mat}}^{\text{particle}}&=&T\sum_{\epsilon=\pm}\sum_{\ell=-\infty}^{+\infty}\int d\tilde{p}
\ln\left(1+e^{-\beta\left(\omega_{p}-\mu_{\epsilon,\ell}\right)}\right),
\nonumber\\
P_{\text{mat}}^{\text{antiparticle}}&=&T\sum_{\epsilon=\pm}\sum_{\ell=-\infty}^{+\infty}\int d\tilde{p}
\ln\left(1+e^{-\beta\left(\omega_{p}+\mu_{\epsilon,\ell}\right)}\right).\nonumber\\
\end{eqnarray}
In the above expressions, the energy-momentum dispersion $\omega_{p}$ and the effective chemical potential $\mu_{\pm,\ell}$ are given by
\begin{eqnarray}\label{N15}
\omega_{p}\equiv \left(\bs{p}_{\perp}^{2}+p_{z}^{2}+m^{2}\right)^{1/2},
\end{eqnarray}
and
\begin{eqnarray}\label{N16}
\mu_{\pm,\ell}\equiv \mu_{\pm}+\ell\Omega, \quad\text{with}\quad \mu_{\pm}\equiv \mu\pm \frac{\Omega}{2}.
\end{eqnarray}
The $+$ and $-$ signs denote the contributions from spin-up ($+$) and spin-down ($-$) particles.
In what follows, we focus on $P_{\text{mat}}^{\text{particle}}$. For simplicity, we omit the sub- and superscripts and denote it by $P$. To perform the summation over $\ell$ in \eqref{N14}, we use the following regularization
\begin{eqnarray}\label{N17}
\sum_{\ell=-\infty}^{+\infty}\ln\left(1+\alpha e^{\beta\ell\Omega}\right)=\ln\left(1+\alpha\right)+\text{divergent terms},\nonumber\\
\end{eqnarray}
for any generic $\ell$-independent factor $\alpha$ [see Appendix \ref{appA} for the proof of \eqref{N17}], and arrive first at
\begin{eqnarray}\label{N18}
P=P_{+}+P_{-},
\end{eqnarray}
with
\begin{eqnarray}\label{N19}
P_{\pm}=T\int d\tilde{p}\ln\left(1+z_{\pm}e^{-\beta\omega_{p}}\right).
\end{eqnarray}
Here, we have introduced the fugacities $z_{\pm}\equiv e^{\beta\mu_{\pm}}$ corresponding to spin-up and spin-down particles. In this regularization scheme, the pressure includes only the spin-rotation coupling through the term $\Omega/2$ in the definition of $\mu_{\pm}$. To determine $P_{\pm}$, we approximate $\omega_{p}$ in two NR and UR limits,
\begin{eqnarray}\label{N20}
\text{NR}:&\qquad&\omega_{p}\simeq \frac{p^{2}}{2m},\nonumber\\
\text{UR}:&\qquad&\omega_{p}\simeq p,
\end{eqnarray}
where $p\equiv |\bs{p}|$.\footnote{The fermion rest mass is neglected in the NR limit.} Plugging $\omega_{p}$ from \eqref{N20} into \eqref{N19} and performing the integration over $p_{\perp}$ and $p_{z}$, we arrive at
\begin{eqnarray}\label{N21}
P_{a,\pm}=\frac{T}{\lambda_{a,T}^{3}}f_{\kappa_{a}+1}(z_{a,\pm}).
\end{eqnarray}
Here, the subscript $a\in\{\text{nr,ur}\}$ corresponds to NR and UR limits, which are characterized by \eqref{N20}. Moreover, $\kappa_{\text{nr}}$ and $\kappa_{\text{ur}}$ are given by $\kappa_{\text{nr}}=3/2$ and $\kappa_{\text{ur}}=3$. For our future purposes, we introduce the fugacity
\begin{eqnarray}\label{N22}
z_{a,\pm}\equiv e^{\beta\mu_{a,\pm}},\quad\mbox{with}\quad \mu_{a,\pm}=\mu_{a}\pm\frac{\Omega_{a}}{2},
\end{eqnarray}
including, in particular, a spin-rotation coupling in $\pm\Omega_{a}/2$ term. The thermal wavelength in these two limits is defined by
\begin{eqnarray}\label{N23}
\lambda_{\text{nr},T}\equiv \left(\frac{2\pi}{mT}\right)^{1/2},\qquad
\lambda_{\text{ur},T}\equiv \left(\frac{\pi^{2}}{T^{3}}\right)^{1/3}.
\end{eqnarray}
Moreover, in \eqref{N21}, the Fermi integral $f_{\nu}(z)$ reads
\begin{eqnarray}\label{N24}
f_{\nu}(z)\equiv\frac{1}{\Gamma(\nu)}\int_{0}^{\infty}\frac{x^{\nu-1}dx}{z^{-1}e^{x}+1}.
\end{eqnarray}
Using the thermodynamic relation
\begin{eqnarray}\label{N25}
n=\left(\frac{\partial P}{\partial\mu}\right)_{T,\Omega},
\end{eqnarray}
it is possible to determine the number densities of spin-up and spin-down particles in the NR and UR limits. They are given by
\begin{eqnarray}\label{N26}
n_{a,\pm}=\frac{1}{\lambda_{a,T}^{3}}f_{\kappa_{a}}(z_{a,\pm}).
\end{eqnarray}
The total number of particles is thus given by $n_{a}=n_{a,+}+n_{a,-}$. To determine $n_{a,\pm}$ at zero temperature, we use
\begin{eqnarray}\label{N27}
f_{\nu}(z_{a,\pm})\simeq\frac{\left(\ln z_{a,\pm}\right)^{\nu}}{\Gamma\left(\nu+1\right)},
\end{eqnarray}
and arrive at
\begin{eqnarray}\label{N28}
n_{a,\pm}=\frac{p_{a,F}^{3}}{6\pi^{2}}\left(1\pm \eta_{a}\right)^{\kappa_{a}},
\end{eqnarray}
with the "spin-chemicorotational ratio" $\eta_{a}\equiv \frac{\Omega_{a}^{(0)}}{2\mu_{a}^{(0)}}$ and the Fermi momentum $p_{a,F}$ defined by
\begin{eqnarray}\label{N29}
p_{nr,F}\equiv \left(2m\mu_{nr}^{(0)}\right)^{1/2}, \qquad p_{ur,F}\equiv \mu_{ur}^{(0)}.
\end{eqnarray}
To determine $\mu_{a}^{(0)}$ and $\Omega_{a}^{(0)}$, let us consider \eqref{N26},
\begin{eqnarray}\label{N30}
n_{a,\pm}\lambda_{a,T}^{3}=f_{\kappa_{a}}(z_{a,\pm}).
\end{eqnarray}
Plugging \eqref{N27} on the right hand side (r.h.s.) of \eqref{N30} and using the definition of $z_{a,\pm}^{(0)}\equiv\exp\left({\beta\mu_{a,\pm}^{(0)}}\right)$, we obtain
\begin{eqnarray}\label{N31}
n_{a,\pm}\simeq \frac{(\beta \mu_{a,\pm}^{(0)})^{\kappa_{a}}}{\lambda_{a,T}^{3}\Gamma\left(\kappa_{a}+1\right)}.
\end{eqnarray}
\begin{figure}
\includegraphics[width=9cm, height=9cm]{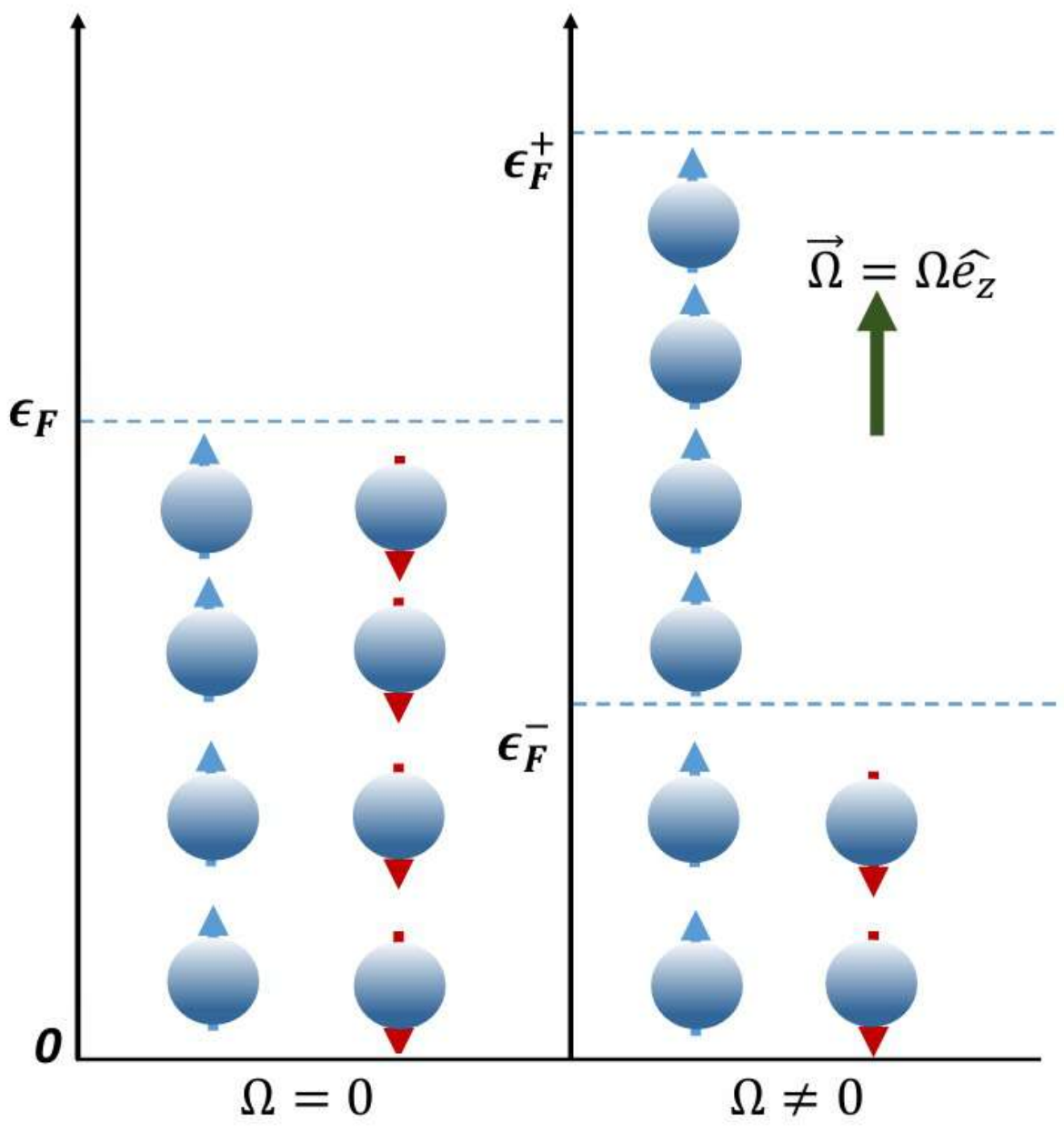}
\caption{The effect of rotation on the splitting of the Fermi energies corresponding to spin-up and spin-down particles at $T=0$ is illustrated. In the left panel, $\epsilon_{F}$ represents the Fermi energy in the absence of rotation ($\Omega=0$), while $\epsilon_{F}^{+}$ and $\epsilon_{F}^{-}$ in the right panel correspond to the Fermi energies of spin-up and spin-down particles for $\Omega\neq 0$. According to \eqref{D6}, the splitting is adjusted by $\eta=\frac{\Omega}{2\mu}$ at $T=0$. The distribution differences of fermions below the Fermi energies for $\Omega\neq 0$ are characterized by $\gamma$, which can be expressed in terms of $\eta$ (as shown in equation \eqref{D2}) or in terms of spin polarization $\mathcal{P}$ (as shown in equation \eqref{D4}). In this example, we choose $\mathcal{P}=0.5$.}\label{fig1}
\end{figure}
\begin{figure*}
\includegraphics[width=8cm, height=6cm]{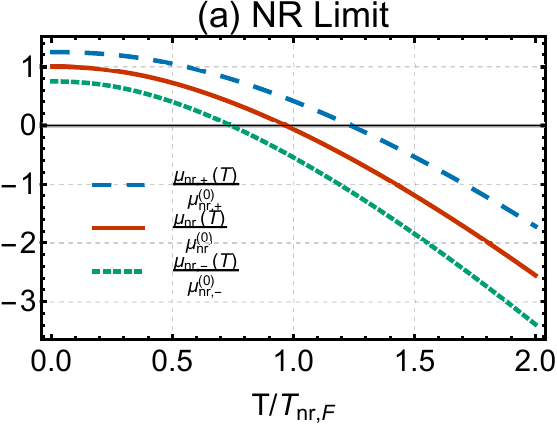}
\includegraphics[width=8cm, height=6cm]{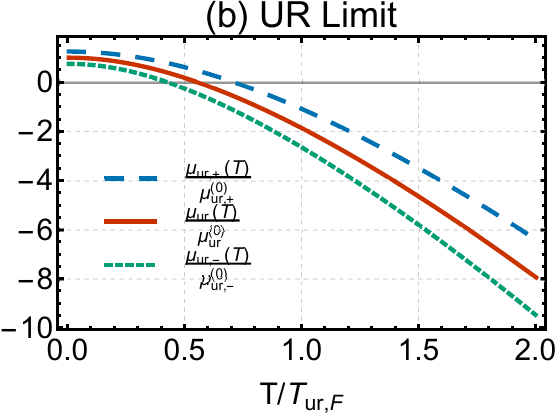}
\caption{The $T/T_{a, F}$ dependence of dimensionless $\frac{\mu_{a,+}}{\mu_{a}^{(0)}}$ (blue dashed curves), $\frac{\mu_{a,-}}{\mu_{a}^{(0)}}$ (green dotted curves), and $\frac{\mu_{a}}{\mu_{a}^{(0)}}$ (red solid curves) are plotted for $a=\text{nr}$ (panel a) and $a=\text{ur}$ (panel b). The numerical results are based on the assumption that $n_{a,\pm}$ remains constant in $T$. It is observed that $\mu_{a,\pm}$ and $\mu_{a}$ decrease as $T$ increases. The temperature dependence of $\mu_{a,\pm}$ is categorized into three distinct temperature regimes, characterized by the signs of $\mu_{a,\pm}$ (see the main text for further details).}\label{fig2}
\end{figure*}
This leads immediately to a relation between $\mu_{a,\pm}^{(0)}$ and $n_{a,\pm}$,
\begin{eqnarray}\label{N32}
\mu_{a,\pm}^{(0)}&\simeq&  T\left(n_{a,\pm}\lambda_{a,T}^{3}\Gamma(\kappa_{a}+1)\right)^{1/\kappa_{a}}\nonumber\\
&=&
\begin{cases}
\text{NR:}&\dfrac{1}{2m}\left(6\pi^{2}n_{nr,\pm}\right)^{2/3}\\
&\\
\text{UR:}&\left(6\pi^{2}n_{ur,\pm}\right)^{1/3}.
\end{cases}
\end{eqnarray}
Here, $\Gamma(5/2)=3\sqrt{\pi}/4$ and $\Gamma(4)=6$ are used.
Combining $\mu_{a,+}^{(0)}$ and $\mu_{a,-}^{(0)}$, we obtain $\mu^{(0)}$ and $\Omega^{(0)}$ of the completely degenerate rotating Fermi gas at $T=0$,
\begin{eqnarray}\label{N33}
\mu_{a}^{(0)}=\frac{1}{2}\left(\mu_{a,+}^{(0)}+\mu_{a,-}^{(0)}\right),\qquad \Omega_{a}^{(0)}=\mu_{a,+}^{(0)}-\mu_{a,-}^{(0)}. \nonumber\\
\end{eqnarray}
We identify $\mu_{a,\pm}^{(0)}$ with the Fermi energies corresponding to spin-up and spin-down fermions, which we denote as $\epsilon_{a,F}^{\pm}$. For $T\ll \epsilon_{a,F}^{\pm}$, both components of the rotating Fermi gas are completely degenerate. In Sec. \ref{sec3A}, we reevaluate the relativistic Barnett effect
in the completely degenerate Fermi gas at zero temperature. We define the spin polarization of the Fermi gas and demonstrate that it is characterized by $\eta_{a}$. In Sec. \ref{sec3B}, we analyze the $T$ dependence of $\mu_{a}$ and $\Omega_{a}$ under the assumption that $n_{a,\pm}$ is independent of $T$. We use $\mu_{a}^{(0)}$ and $\Omega_{a}^{(0)}$ from \eqref{N33} as the initial values for $\mu_{a}$ and $\Omega_{a}$ to solve the differential equation that arises from this assumption.
\section{Thermodynamic behavior of rotating Fermi gas}\label{sec3}
\subsection{Relativistic Barnett effect in a completely degenerate Fermi gas}\label{sec3A}
\setcounter{equation}{0}
\begin{figure*}
\includegraphics[width=8cm, height=6cm]{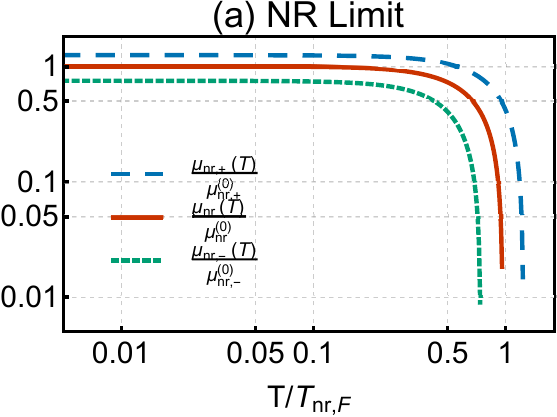}
\includegraphics[width=8cm, height=6cm]{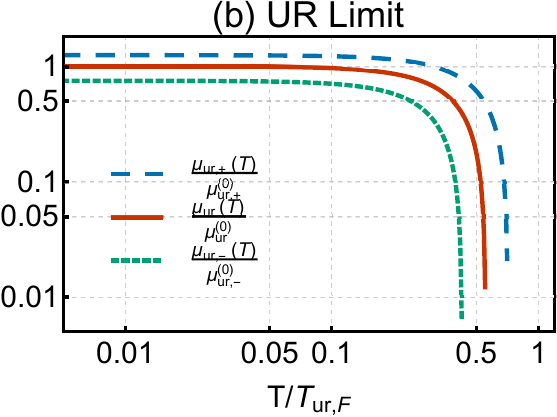}
\caption{The log-log plots of the $T/T_{a, F}$ dependence of dimensionless $\frac{\mu_{a,+}}{\mu_{a}^{(0)}}$ (blue dashed curves), $\frac{\mu_{a,-}}{\mu_{a}^{(0)}}$ (green dotted curves), and $\frac{\mu_{a}}{\mu_{a}^{(0)}}$ (red solid curves) are presented for $a=\text{nr}$ (panel a) and $a=\text{ur}$ (panel b). It turns out that in both NR and UR limits, the spin-down component of the rotating Fermi gas dilutes at a lower temperature compared to its spin-up counterpart.  }\label{fig3}
\end{figure*}
To describe the impact of rigid rotation on spin polarization in a completely degenerate Fermi gas at $T=0$, we consider a fermion gas with $n_{a,+}$ spin-up and $n_{a,-}$spin-down fermions, in a unit volume. We assume that in the absence of rotation, $n_{a,+}=n_{a,-}$. When rotation is initiated with angular velocity $\bs{\Omega}$, a significant number of particles align their spins in the direction of $\Omega$ to minimize the potential energy, which is given by $U_{\Omega}\equiv -\bs{S}\cdot \bs{\Omega}$. To characterize the ratio $\frac{n_{a,+}}{n_{a,-}}$ in the rotating system, we introduce a parameter $\mathcal{P}_{a}$,
\begin{eqnarray}\label{D1}
\gamma_{a}\equiv \frac{n_{a,+}}{n_{a,-}}=\frac{1+\mathcal{P}_{a}}{1-\mathcal{P}_{a}}.
\end{eqnarray}
Here, $\mathcal{P}_{a}$ is defined in the expression $n_{a,\pm}=\frac{1}{2}(1\pm\mathcal{P}_{a})n_{a}$.
Using \eqref{N31} and the definition of $\mu_{a,\pm}^{(0)}$ in terms of $\mu_{a}^{(0)}$ and $\Omega_{a}^{(0)}$, $\gamma_{a}$ is given by
\begin{eqnarray}\label{D2}
\gamma_{a}=\left(\frac{\mu_{a,+}^{(0)}}{\mu_{a,-}^{(0)}}\right)^{\kappa_{a}}=\left(\frac{1+\eta_{a}}{1-\eta_{a}}\right)^{\kappa_{a}},\quad\mbox{with}\quad \eta_{a}\equiv \frac{\Omega_{a}^{(0)}}{2\mu_{a}^{(0)}}. \nonumber\\
\end{eqnarray}
Reformulating \eqref{D1}, it turns out that $\mathcal{P}_{a}$ plays the role of the spin polarization of the Fermi gas, defined in the literature (see, e.g., \cite{sahoo2025b}),
\begin{eqnarray}\label{D3}
\mathcal{P}_{a}=\frac{n_{a,+}-n_{a,-}}{n_{a,+}+n_{a,-}}.
\end{eqnarray}
This factor can also be interpreted as the net spin (number) density normalized by the total spin (number) density. On the other hand, \eqref{D1} yields
\begin{eqnarray}\label{D4}
\mathcal{P}_{a}=\frac{\gamma_{a}-1}{\gamma_{a}+1},
\end{eqnarray}
with $\gamma_{a}$ from \eqref{D2}. Plugging $\gamma_{a}$ from \eqref{D2} into \eqref{D4}, we arrive at $\eta_{a}$ in terms of $\mathcal{P}_{a}$,
\begin{eqnarray}\label{D5}
\eta_{a}=\frac{\gamma_{a}^{1/\kappa_{a}}-1}{\gamma_{a}^{1/\kappa_{a}}+1},
\end{eqnarray}
with $\gamma_{a}$ from \eqref{D1}. For $\mathcal{P}_{a}=0.5$, we obtain $\eta_{nr}=0.35$ and $\eta_{ur}=0.18$.
\par
The polarization $\mathcal{P}_{a}$ satisfies $0<\mathcal{P}_{a}<1$. To show this, we note that when $\eta_{a}<1$, $\mathcal{P}_{a}$ is positive. This occurs because the onset of rotation in a given direction with angular velocity $\bs{\Omega}$, causes a large number of fermions to align their spins with $\bs{\Omega}$ in order to minimize the energy. As a result, we find that  $n_{a,-}<n_{a,+}$, which implies $\mathcal{P}_{a}>0$. Furthermore, as long as the condition $\Omega_{a}^{(0)}<2\mu_{a}^{(0)}$ is satisfied, $n_{a,-}<n_{a,+}$. This leads to the conclusion that $n_{a,+}<n_{a}$, necessiating that $\mathcal{P}_{a}<1$.
\par
To understand the effect of rotation, we identify $\mu_{a,\pm}^{(0)}$ at $T=0$ with the Fermi energies corresponding to the spin-up and spin-down particles, denoted as $\epsilon_{F}^{\pm}$. According to \eqref{N16}, these energies are given by
\begin{eqnarray}\label{D6}
\epsilon_{a,F}^{\pm}\equiv \mu_{a,\pm}^{(0)}=\mu_{a}^{(0)}\pm\frac{\Omega_{a}^{(0)}}{2}=\mu_{a}^{(0)}\left(1\pm \eta_{a}\right).
\end{eqnarray}
For $\Omega_{a}^{(0)}=0$, $\epsilon_{a,F}^{\pm}= \mu_{a}^{(0)}$ for both spin-up and spin-down particles. When rotation begins, the Fermi energies $\epsilon_{F}^{\pm}$ become split. Assuming $\Omega_{a}^{(0)}>0$, we find $\epsilon_{F}^{-}<\epsilon_{F}^{+}$, which directly leads to the conclusion that $n_{a,-}<n_{a,+}$. In this scenario, $\eta_{a}$ adjusts the value of $\mathcal{P}_{a}$ and vice versa. Figure \ref{fig1} illustrates the effect of $\Omega$ on the splitting of the Fermi energies for $n_{a}=8$ fermions per unit volume. For $\Omega^{(0)}=0$, both $n_{a,+}$ and $n_{a,-}$ equal $4$ (left panel). Assuming $\mathcal{P}_{a}=0.5$ and applying \eqref{D1}, we obtain $\gamma_{a}=3$ as rotation is initiated. This results in $n_{a,+}=6$ and $n_{a,-}=2$ (right panel). As expected, $n_{a,+}>n_{a,-}$. Furthermore, plugging $\mathcal{P}_{a}$ into \eqref{D5} and using $\kappa_{\text{nr}}=3/2$ and $\kappa_{\text{ur}}=3$, we find $\eta_{\text{nr}}\sim 0.35$ and $\eta_{\text{ur}}\sim 0.18$. The Fermi energies of spin-up and spin-down fermions are then given by \eqref{D6}.
Notably, the fact that the spin-chemicorotational ratio $\eta_{a}$ of a rotating Fermi gas adjusts the spin polarization $\mathcal{P}_{a}$ is a novel finding. This may have implications for the physics of QGP, where research on issues related to $\Lambda$ polarization is an active field of study.
\subsection{$\bs{T}$ dependence of $\bs{\mu_{a,\pm}}$ and $\bs{\Omega_{a,\pm}}$}\label{sec3B}
\begin{figure*}
\includegraphics[width=8cm, height=6cm]{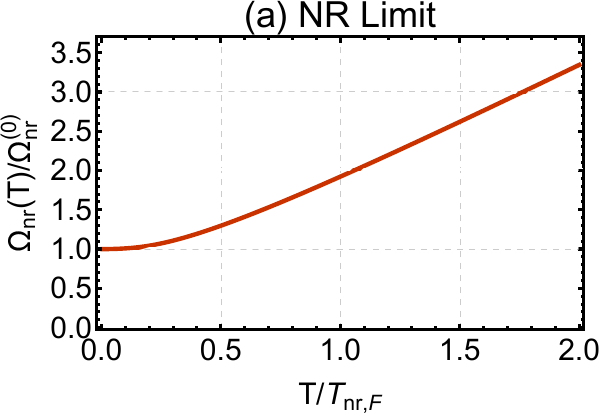}
\includegraphics[width=8cm, height=6cm]{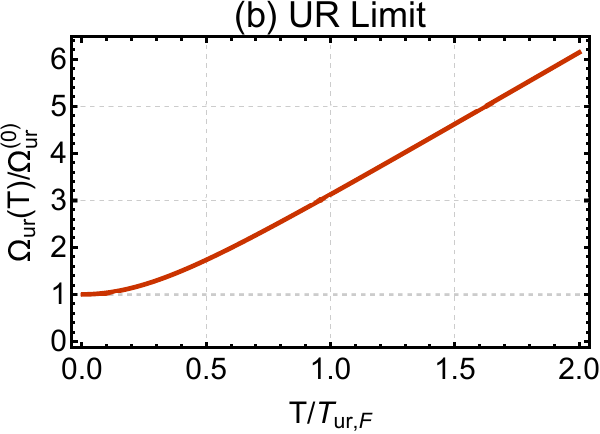}
\caption{The $T/T_{a, F}$ dependence of dimensionless $\Omega_{a}/\Omega_{a}^{(0)}$ are plotted for $a=\text{nr}$ (panel a) and $a=\text{ur}$ (panel b).  The numerical results are based on the assumption that $n_{a,\pm}$ remains constant as $T$ increases. It is observed that $\Omega_{a}$ increases as $T$ rises.}\label{fig4}
\end{figure*}
After determining the spin polarization $\mathcal{P}_{a}$ at $T=0$, we assume that $n_{a,\pm}$ and consequently $\mathcal{P}_{a}$ remain constant as  $T$ increases. In other words, we have
\begin{eqnarray}\label{D7}
\frac{\partial n_{a,\pm}}{\partial T}=0.
\end{eqnarray}
Plugging $n_{a,\pm}$ from \eqref{N26} into \eqref{D7}, we arrive at a differential equation for the fugacity $z_{a,\pm}$ from \eqref{N22},
\begin{eqnarray}\label{D8}
\frac{1}{z_{a,\pm}}\left(\frac{\partial z_{a,\pm}}{\partial T}\right)_{n_{a,\pm}}=-\frac{\kappa_{a}}{T}\frac{f_{\kappa_{a}}(z_{a,\pm})}{f_{\kappa_{a}-1}(z_{a,\pm})}.
\end{eqnarray}
We have solved this equation numerically by choosing fixed values for $\mu_{a}^{(0)}$ and $\Omega_{a}^{(0)}$ at $T=0$. In Fig. \ref{fig2}, the $T$ dependence of normalized $\mu_{a,+}/\mu_{a}^{(0)}$ (blue dashed curves), $\mu_{a,-}/\mu_{a}^{(0)}$ (green dotted curves), and $\mu_{a}/\mu_{a}^{(0)}$ (red solid curves) is demonstrated for the NR limit  ($a=\text{nr}$) [Fig. \ref{fig2}(a)] and UR limit ($a=\text{ur}$)  [Fig. \ref{fig2}(b)]. Here, $\mu_{a}\equiv \left(\mu_{a,+}+\mu_{a,-}\right)/2$ [see \eqref{N33}]. For these plots, we have chosen $\mu_{a}^{(0)}=2$ MeV,
$\Omega_{a}^{(0)}=1$ MeV. The Fermi temperature $T_{a,F}$ is defined as $T_{a,F}=\mu_{a}^{(0)}$.\footnote{We note that this choice of parameters is generic and not specific to any particular phenomenological example. We present only dimensionless quantities in Figs. \ref{fig2}-\ref{fig5}.}
\par
According to the results presented in Fig. \ref{fig2}, we can identify three distinct temperature regimes, characterized by the signs of $\mu_{a,\pm}$:
\begin{enumerate}
\item[\textit{i)}] \textit{Low-temperature regime}, where $\mu_{a,+}>0$ and $\mu_{a,-}>0$,
\item[\textit{ii)}] \textit{Intermediate-temperature regime}, where $\mu_{a,+}>0$ and $\mu_{a,-}<0$,
\item[\textit{iii)}] \textit{High-temperature regime}, where $\mu_{a,+}<0$ and $\mu_{a,-}<0$.
\end{enumerate}
Numerically, the low-temperature regime occurs at $T< 0.73 T_{\text{nr},F}$ for the NR limit and $T<0.44 T_{\text{ur},F}$ for the UR limit, the intermediate-temperature regime is given by $0.73 T_{\text{nr},F}<T<1.22 T_{\text{nr},F}$ for the NR limit and $0.44 T_{\text{ur},F}<T<0.73 T_{\text{ur},F}$ for the UR limit, and the low-temperature regime occurs at $T>1.22 T_{\text{nr},F}$ for the NR limit and $T>0.73 T_{\text{ur},F}$ for the UR limit. The two components of the rotating Fermi gas display different behavior in these regimes. In the low-temperature regime, both components exhibit strong degeneracy. In the intermediate-temperature regime, spin-up fermions remain strongly degenerate, while spin-down fermions become weakly degenerate. In the high-temperature regime, both components of the gas are weakly degenerate. The distinction between these regimes allows us to analytically determine the $T$ dependence of $\mu_{a}$.
It is noteworthy that in the absence of rotation, the Fermi gas consists of only one component, which is either in a strongly or weakly degenerate regime \cite{rebhan-book}.
\par
In what follows, we utilize the asymptotic approximation of $f_{\kappa_{a}}(z_{a,\pm})$ in \eqref{N30} for $z_{a,\pm}\gg 1$ (strongly degenerate gas at low temperature) and $z_{a,\pm}\ll 1$ (weakly degenerate gas at high temperature). Our goal is to determine the $T$ dependence of $\mu_{a}$ in these regimes.
\par
\textit{i) Low-temperature regime:} In this regime, both components of the Fermi gas are at low temperatures, and their corresponding $\mu_{a,\pm}$ are positive. In the previous section, we determined $\mu_{a,\pm}^{(0)}$ at $T=0$ by making use of \eqref{N27}. It is given by \eqref{N32}. To determine the $T$ dependence of $\mu_{a,\pm}$ in the low-temperature regime, we need the next-to-leading term in \eqref{N27}. This term is provided by the second term of the Sommerfeld expansion,
\begin{eqnarray}\label{D9}
f_{\nu}(z)\simeq\frac{(\ln z)^{\nu}}{\Gamma(\nu+1)}\left(1+\frac{\pi^{2}}{6}\nu(\nu-1)\left(\ln z\right)^{-2}\right).\nonumber\\
\end{eqnarray}
Plugging \ref{D9}, into \eqref{N30}, we arrive first at
\begin{eqnarray}\label{D10}
\lefteqn{\hspace{-0.5cm}n_{a,\pm}\lambda_{a,T}^{3}\simeq \frac{\left(\ln z_{a,\pm}\right)^{\kappa_{a}}}{\Gamma\left(\kappa_{a}+1\right)}
}
\nonumber\\
&&\times\left(1+\frac{\pi^{2}}{6}\kappa_{a}\left(\kappa_{a}-1\right)\left(\ln z_{a,\pm}^{(0)}\right)^{-2}\right). 
\end{eqnarray}
Then, using the zeroth order correction \eqref{N31}, we obtain
\begin{eqnarray}\label{D11}
\hspace{-0.5cm}\mu_{a,\pm}\simeq T_{a,F}^{\pm}\left(1-\frac{\pi^{2}}{6}(\kappa_{a}-1)\left(\frac{T}{T_{a,F}^{\pm}}\right)^{2}\right).
\end{eqnarray}
Here, we introduced the Fermi temperature $T_{a,F}^{\pm}\equiv \mu_{a,\pm}^{(0)}$. According to \ref{D11}, $\mu_{a,\pm}$ is positive for $T\gtrsim \mathcal{T}_{a,\pm} $ with
\begin{eqnarray}\label{D12}
\mathcal{T}_{a,\pm}\equiv \frac{T_{a,F}^{\pm}}{[\pi^{2}\left(\kappa_{a}-1\right)/6]^{1/2}}.
\end{eqnarray}
\textit{ii) Intermediate-temperature regime:} In this regime $\mu_{a,+}>0$ and $\mu_{a,-}<0$.  Using \eqref{D11}, it turns out that
\begin{eqnarray}\label{D13}
\hspace{-0.5cm}\mu_{a,+}\simeq T_{a,F}^{+}\left(1-\frac{\pi^{2}}{6}(\kappa_{a}-1)\left(\frac{T}{T_{a,F}^{+}}\right)^{2}\right).
\end{eqnarray}
For $T\lesssim \mathcal{T}_{a,+}$, we have  $\mu_{a,+}>0$.
As concerns $\mu_{a,-}$, however, we use the high-temperature (small $z_{a}$) expansion of $f_{\nu}(z_{a})$,
\begin{eqnarray}\label{D14}
f_{\nu}(z_{a})=z_{a}-\frac{z_{a}^{2}}{2^{\nu}}+\frac{z_{a}^{3}}{3^\nu}+\cdots.
\end{eqnarray}
We consider only the first two terms and arrive at
\begin{eqnarray}\label{D15}
z_{a}\approx f_{\nu}(z_{a})+\frac{(f_{\nu}(z_{a}))^{2}}{2^{\nu}}.
\end{eqnarray}
Plugging \eqref{D15} into \eqref{N30}, we obtain
\begin{eqnarray}\label{D16}
\mu_{a,-}(T)\approx T\ln\left(\zeta_{a,-}+\frac{\zeta_{a,-}^{2}}{2^{\kappa_{a}}}\right),
\end{eqnarray}
with
\begin{eqnarray}\label{D17}
\zeta_{a,\pm}(T)\equiv \frac{1}{\Gamma(\kappa_{a}+1)}\left(\frac{T_{a,F}^{\pm}}{T}\right)^{\kappa_{a}}.
\end{eqnarray}
For $T\gtrsim\tilde{\mathcal{T}}_{a,-}$, with $\tilde{\mathcal{T}}_{a,-}$ arising from the solution of
\begin{eqnarray}\label{D18}
\zeta_{a,-}(T)+\frac{\zeta_{a,-}^{2}(T)}{2^{\kappa_{a}}}=1,
\end{eqnarray}
we have $\mu_{a,-}<0$.
\par
\textit{iii) High-temperature regime:} This regime is characterized with $\mu_{a,+}<0$ and $\mu_{a,-}<0$. Both component of gases are weakly degenerate. To determine $\mu_{a,\pm}$ in this regime, we use \eqref{D14}. Similar to \eqref{D16}, we arrive at
 \begin{eqnarray}\label{D19}
\mu_{a,\pm}\approx T\ln\left(\zeta_{a,\pm}+\frac{\zeta_{a,\pm}^{2}}{2^{\kappa_{a}}}\right),
\end{eqnarray}
with $\zeta_{a,\pm}$ from \eqref{D17}. For $T\gtrsim \bar{\mathcal{T}}_{a,\pm}$, with $\bar{\mathcal{T}}_{a,\pm}$ arising from the solution of
\begin{eqnarray}\label{D20}
\zeta_{a,\pm}(T)+\frac{\zeta_{a,\pm}^{2}(T)}{2^{\kappa_{a}}}=1,
\end{eqnarray}
we have $\mu_{a,\pm}<0$. We note that in all three regimes, the relation
$\mu_{a}=(\mu_{a,+}+\mu_{a,-})/2$ and $\Omega_{a}=\mu_{a,+}-\mu_{a,-}$ remain valid, leading to the desired $T$ dependence of $\mu_{a}$ and $\Omega_{a}$. To compare our results for $\mu_{a}$ in a nonrotating Fermi gas \cite{rebhan-book}, log-log plots of dimensionless $\mu_{a,\pm}(T)/\mu_{a,\pm}^{(0)}$ and $\mu_{a}(T)/\mu_{a}^{(0)}$ as a function of $T/T_{a,F}$ are plotted in Fig. \ref{fig3}. Our findings indicate that in both NR and UR limits, the spin-down component of the rotating Fermi gas becomes weakly degenerate at lower temperatures than its spin-up component. Additionally, the ultrarelativistic gas becomes less dense at lower temperatures than the nonrelativistic Fermi gas.
\par
In Fig. \ref{fig4}, the numerical results for $z_{a,\pm}$ are used and the $T$ dependence of dimensionless $\Omega_{a}(T)/\Omega_{a}^{(0)}$ is plotted. As it is shown, $\Omega_{a}$ increases with rising temperature. It is important to note that the $T$ dependence of $\Omega_{a}$ is based on the assumption that the number density of spin-up and spin-down components of the rotating Fermi gas remain temperature independent. 
\subsection{Curie law for the moment of inertia}\label{sec3C}
\begin{figure*}
\includegraphics[width=8cm, height=6cm]{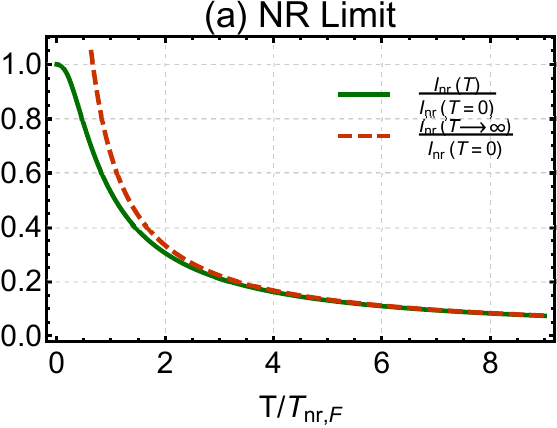}
\includegraphics[width=8cm, height=6cm]{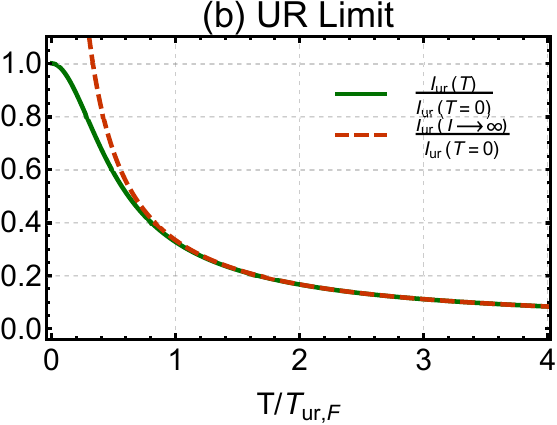}
\caption{The $T/T_{a, F}$ dependence of the ratio $I_{a}(T)/I_{a}(T=0)$ (solid curves) and $I_{a}(T\to \infty)/I_{a}(T=0)$ (dashed curves) are plotted for $a=\text{nr}$ (panel a) and $a=\text{ur}$ (panel b).  The asymptotic behavior of the moment of inertia for large $T$ exhibits a $1/T$ law [see \eqref{D30}]. This behavior is similar to the Curie law for the magnetic susceptibility $\chi_{m}$, as discussed in the main text.}\label{fig5}
\end{figure*}
Using the data for $z_{a,\pm}$ obtained from the numerical solution of \eqref{D8}, we can determine several thermodynamic quantities. In this section, we focus on the moment of inertia and show that, at high temperature, it is proportional to $1/T$. This behavior is analogous to the behavior of the magnetic susceptibility and is referred to as the Curie law of paramagnetism.
\par
The moment of inertia is defined
\begin{eqnarray}\label{D21}
I\equiv \frac{\partial J}{\partial\Omega},
\end{eqnarray}
where $J$ is the angular momentum density given by
\begin{eqnarray}\label{D22}
J\equiv \left(\frac{\partial P}{\partial\Omega}\right)_{T,\mu}.
\end{eqnarray}
Here, $P$ is the total pressure of the rotating Fermi gas. Because of different spin fugacities of the two components of the rotating Fermi gas, the total angular momentum density and moment of inertia are given by $J_{a}=J_{a,+}+J_{a,-}$ and $I_{a}=I_{a,+}+I_{a,-}$ with $J_{a,\pm}$ and $I_{a,\pm}$ corresponding to spin-up and spin-down fermions. Plugging the pressure $P_{a,\pm}$ from \eqref{N21} into \eqref{D22} and \eqref{D21}, they are given by
\begin{eqnarray}\label{D23}
J_{a,\pm}&=&\pm\frac{1}{2\lambda_{a,T}^{3}}f_{\kappa_{a}}\left(z_{a,\pm}\right),\nonumber\\
I_{a,\pm}&=&\frac{1}{4T\lambda_{a,T}^{3}}f_{\kappa_{a}-1}\left(z_{a,\pm}\right).
\end{eqnarray}
Using \eqref{N27}, $I_{a,\pm}$ at zero temperature is given by
\begin{eqnarray}\label{D24}
I_{a,\pm}(T=0,\Omega_{a}^{(0)})\simeq \frac{\left(\beta\mu_{a,\pm}^{(0)}\right)^{\kappa_{a}-1}}{\lambda_{a,T}^{3}\Gamma\left(\kappa_{a}\right)}.
\end{eqnarray}
Moreover, utilizing \eqref{N31}, the ratio $I_{a,\pm}/n_{a,\pm}$ at $T=0$ reads
\begin{eqnarray}\label{D25}
\frac{I_{a,\pm}(T=0,\Omega_{a}^{(0)})}{n_{a,\pm}(T=0,\Omega_{a}^{(0)})}=\frac{\kappa_{a}}{4\mu_{a,\pm}^{(0)}}.
\end{eqnarray}
Plugging
\begin{eqnarray}\label{D26}
\hspace{-0.5cm}\mu_{a,\pm}^{(0)}=T_{a, F}\left(1\pm \eta_{a}\right), \quad \text{with}\quad \eta_{a}=\frac{\Omega_{a}^{(0)}}{2\mu_{a}^{(0)}},
\end{eqnarray}
and $T_{a,F}\equiv\mu_{a}^{(0)}$ into \eqref{D25}, we obtain
\begin{eqnarray}\label{D27}
\frac{I_{a,\pm}(T=0,\Omega_{a}^{(0)})}{n_{a,\pm}(T=0,\Omega_{a}^{(0)})}=\frac{\kappa_{a}}{4 T_{a,F}\left(1\pm \eta_{a}\right)}.
\end{eqnarray}
At this stage, we set $\Omega_{a}^{(0)}=0$ to consider only the linear response to rotation. We arrive at
\begin{eqnarray}\label{D28}
\bar{I}_{a,\pm}=\frac{\kappa_{a}\bar{n}_{a,\pm}}{4 T_{a,F}},
\end{eqnarray}
where $\bar{I}_{a,\pm}\equiv I_{a,\pm}(T=0,\Omega_{a}^{(0)}=0)$ and $\bar{n}_{a,\pm}\equiv n_{a,\pm}(T=0,\Omega_{a}^{(0)}=0)$.
\par
On the other hand, at high-temperature limit, we use \eqref{D14} and replace $f_{\kappa_{a}}(z_{a,\pm})$ in \eqref{N26} and $f_{\kappa_{a}-1}(z_{a,\pm})$ in \eqref{D23} with $z_{a,\pm}$. We obtain
\begin{eqnarray}\label{D29}
\tilde{I}_{a,\pm}=\frac{\tilde{n}_{a,\pm}}{4T}.
\end{eqnarray}
Here, $\tilde{I}_{a,\pm}\equiv I_{a,\pm}(T\to \infty,\Omega_{a}^{(0)}=0)$ and $\tilde{n}_{a,\pm}\equiv n_{a,\pm}(T\to \infty,\Omega_{a}^{(0)}=0)$. The ratio $\tilde{I}_{a}/\bar{I}_{a}$ is thus given by
\begin{eqnarray}\label{D30}
\frac{\tilde{I}_{a}}{\bar{I}_{a}}=\frac{1}{\kappa_{a}}\frac{T_{a,F}}{T}.
\end{eqnarray}
In this analysis, we have assumed that the number density at $T=0$ does not change with $T$ and have set
$\tilde{n}_{a}=\bar{n}_{a}$.
In Fig. \ref{fig5}, the $T/T_{a,F}$ dependence of the ratio $I_{a}/\bar{I}_{a}$ with $I_{a}=I_{a,+}+I_{a,-}$ and $\bar{I}_{a}=\bar{I}_{a,+}+\bar{I}_{a,-}$ is plotted (green solid curves) for $a=\text{NR}$ [Fig. \ref{fig5}(a)] and $a=\text{UR}$ [Fig. \ref{fig5}(b)]. The green solid curves are determined by choosing $\Omega_{a}^{(0)}=0$ as the initial value of $\Omega_{a}$ and solving \eqref{D8} numerically to determine the corresponding $z_{a,\pm}$. Plugging these data into \eqref{D23}, we arrived at $I_{a,\pm}(\Omega=0)$.
According to the results in Figs. \ref{fig5}(a) and \ref{fig5}(b), in both NR and UR limits, $I_{a}$ is positive and decreases with increasing $T$. 
The red dashed curves in Fig. \ref{fig5} demonstrate the ratio $\tilde{I}_{a}/\bar{I}_{a}$ from \eqref{D30}. This ratio is proportional to $1/T$, exhibiting behavior akin to the magnetic susceptibility $\chi_{m}$ at high temperature, as described by the Curie law. In the following discussion, we will explain why this result is expected, drawing an analogy between the Barnett magnetization and angular momentum density. According to this analogy, the moment of inertia plays a role similar to that of the magnetic susceptibility, acting as a linear response to an effective magnetic field described by the angular velocity $\Omega$.
\par
As it is discussed in \cite{sahu2026}, the Barnett magnetization arises from the equivalence between the energy of the magnetic moment in a magnetic field $U_{B}$ and the energy of a particle with spin $\bs{S}$ in a rotating frame $U_{\Omega}$. They are given by
\begin{eqnarray}\label{D31}
U_{B}=-g\mu_{B}\bs{S}\cdot \bs{B},\qquad U_{\Omega}=-\bs{S}\cdot\bs{\Omega}.
\end{eqnarray}
Here, $g$ is the Land$\acute{\text{e}}$ factor and $\mu_{B}=\frac{-e}{2m_{e}}$ is the Bohr magneton. Equating $U_{B}$ and $U_{\Omega}$ an effective magnet field is defined in term of the angular velocity $\Omega$,
\begin{eqnarray}\label{D32}
B_{\text{eff}}\equiv \frac{\Omega}{g\mu_{B}}.
\end{eqnarray}
The Barnett magnetization is defined by
\begin{eqnarray}\label{D33}
M_{\text{Barnett}}\equiv \left(\frac{\partial P}{\partial B_{\text{eff}}}\right)_{T,\mu}\bigg|_{B_{\text{eff}}=0}.
\end{eqnarray}
Plugging $B_{\text{eff}}$ from \eqref{D32} into \eqref{D33} and using \eqref{D22}, we arrive at
\begin{eqnarray}\label{D34}
M_{\text{Barnett}}=g\mu_{B}J.
\end{eqnarray}
Equating, at this stage
\begin{eqnarray}\label{D35}
M_{\text{Barnett}}=\chi_{m}B_{\text{eff}},
\end{eqnarray}
with \eqref{D34} and plugging
\begin{eqnarray}\label{D36}
J=I\Omega,
\end{eqnarray}
from \eqref{D21} for $\Omega=0$ as well as $B_{\text{eff}}$ from \eqref{D32} into the resulting expression, we arrive at
\begin{eqnarray}\label{D37}
\chi_{m}=(g\mu_{B})^{2} I.
\end{eqnarray}
Hence, for $T$ independent proportionality factor $(g\mu_{B})^{2}$, the $T$ dependence of $I$ is the same as that of $\chi_{m}$. The latter follows the Curie law of paramagnetism in nonrotating Fermi gas, and we have identified a similar behavior in a rigidly rotating one in this section.
\section{Conclusions}\label{sec4}
\setcounter{equation}{0}
In this work, we investigated the relativistic Barnett effect in a rigidly rotating Fermi gas. We utilized the Lagrangian density of rigidly rotating free fermions. Following the standard imaginary-time formalism of thermal field theory, we determined the pressure of this medium. As expected, we found that the chemical potential $\mu$ is modified by the angular velocity $\Omega$ expressed as $\mu_{\pm,\ell}=\mu+(\ell\pm 1/2)\Omega$. Here, $\ell$ is the quantum number corresponding to the $z$ component of orbital angular momentum, and $\pm 1/2$ represent the spins of spin-up ($+$) and spin-down ($-$) fermions. Focusing on the thermal part of the pressure and employing a specific regularization scheme (details can be found in Appendix \ref{appA}), we performed a summation over $\ell$. We demonstrated that, within this scheme, the pressure and all thermodynamic quantities derived from it can be separated into two parts, corresponding to spin-up and spin-down fermions. Thus, the angular momentum-dependent parts of these quantities are summed, allowing the thermodynamics to be described solely by their spin-dependent parts. It is important to note that the factor $\pm\Omega/2$ in $\mu_{\pm,\ell}$ represents the spin-rotation coupling. This coupling appears in the spin fugacity of spin-up and spin-down fermions, given by $e^{\pm\beta\Omega/2}$, where $\beta$ denotes the inverse temperature.
\par
In Secs. \ref{sec2} and \ref{sec3}, we determined the pressure, number density, angular momentum density, and moment of inertia of the rotating Fermi gas in two nonrelativistic and ultrarelativistic limits. We expressed these quantities in terms of the Fermi integral \eqref{N24}. By utilizing the asymptotic formula for this function at $T\to 0$, we established a relation between the number density of spin-up and spin-down particles and their corresponding chemical potentials in the completely degenerate Fermi gas at $T=0$. To describe the Barnett effect, we introduced the ratio of the number density of spin-up and spin-down fermions in terms of a parameter $\mathcal{P}_{a}$, which is shown to be the spin polarization of the rotating Fermi gas. We demonstrated that $\mathcal{P}_{a}$ is directly related to the spin-chemicorotational ratio $\eta_{a}=\Omega_{a}^{(0)}/2\mu_{a}^{(0)}$ of this medium, where $\Omega_{a}^{(0)}$ and $\mu_{a}^{(0)}$ are the angular velocity and chemical potential of the completely degenerate rotating Fermi gas at $T=0$. We further showed that for a given value of $\eta_{a}$, the Fermi energies, $\epsilon_{F}^{\pm}$, corresponding to spin-up and spin-down fermions become split. Specifically, when $\eta_{a}>0$, $\epsilon_{F}^{+}$ is greater than $\epsilon_{F}^{-}$. As a result, in a rotating medium, there are more fermions whose spins align with the axis of angular velocity \(\bs{\Omega}\) than those whose spins oppose it. This mechanism leads to the spin polarization characteristic of the Barnett effect. We illustrated this mechanism in Fig. \ref{fig1}.
\par
To determine the $T$ dependence of $\mu_{a}$ and $\Omega_{a}$, we assumed that the number densities of spin-up and spin-down particles remain constant with temperature. We numerically solved the resulting differential equation \eqref{D8} to obtain the corresponding fugacities and demonstrated that $\mu_{a}$ decreases while $\Omega_{a}$ increases the temperature rises. Referring to the data shown in Fig. \ref{fig2}, we identified three different temperature regimes based on the sign of $\mu_{a,\pm}$. In the low-temperature regime, $\mu_{a,+}$ and $\mu_{a,-}$ are positive. In the intermediate-temperature regime $\mu_{a,+}$ is positive while $\mu_{a,-}$ is negative. In the high-temperature regime, both $\mu_{a,+}$ and $\mu_{a,-}$ are negative. Each of these regimes reveals different behaviors for the two components of the rotating Fermi gas. In the low- (high-) temperature regime, both components are strongly (weakly) degenerate. In the intermediate-temperature regime,  the spin-up component remains strongly degenerate, while the spin-down component becomes weakly degenerate. 
Following this reasoning and using the high- and low-temperature expansions of the Fermi integral, we derived the analytical expression for the $T$ dependence of $\mu_{a}$ across these three regimes. Additionally, we found that in both NR and UR limits, the spin-down component of the rotating Fermi gas becomes less dense at lower temperatures compared to the spin-up component.
We emphasize that in the absence of rotation, the Fermi gas consists of only one component, which can either be in the strongly degenerate or in the weakly degenerate regime \cite{rebhan-book}. Consequently, intermediate-temperature regimes arise only when $\Omega$ is nonzero. Regarding the $T$ dependence of $\Omega_{a}$, it increases as temperature rises. To the best of our knowledge, this is the first instance in the literature where the temperature dependence of $\Omega$ is addressed. The rate at which $\Omega$ depends on temperature could have implications for the physics of heavy-ion collisions.
\par
In Sec. \ref{sec3C}, we defined the effective magnetic field $B_{\text{eff}}$ induced by rotation with angular velocity $\Omega$, and showed that the Barnett magnetization $M_{\text{Barnett}}$, resulting from $B_{\text{eff}}$ can be expressed in terms of the angular momentum density $J$. Thus, under a linear approximation for $B_{\text{eff}}$ and $\Omega$, the magnetic susceptibility $\chi_{m}$ associated with $B_{\text{eff}}$ is proportional to the moment of inertia $I$ corresponding to $\Omega$. Based on the Curie law of paramagnetism, we expect that in the high-temperature limit, $I$ behaves as $1/T$. 
In Fig. \ref{fig5}, we plotted the $T$ dependence of $I$ and demonstrated its $1/T$ dependence in the high-temperature regime. This behavior is also analytically confirmed in \eqref{D30}. 
\par
This work can be extended to the case when the rotating Fermi gas is subjected to a homogeneous magnetic field. However, it remains unclear whether the separation of spins, crucial for the discussion on the Barnett effect in this paper, is possible in the presence of an external magnetic field. Additionally, it is important to investigate the effects of dimensional reduction that arises from a homogeneous magnetic field. Additionally, it is crucial to explore the implications of the results presented in this paper on the dynamics of the QGP produced at RHIC and LHC, as this remains an essential and open question.
\begin{appendix}
\section{The proof of \eqref{N17}}\label{appA}
\setcounter{equation}{0}
In this appendix, we prove \eqref{N17}. We start with regularizing the summation over $\ell$ in
\begin{eqnarray}\label{appA1}
\sum_{\ell=-\infty}^{\infty}1, \quad \sum_{\ell=-\infty}^{\infty}\ell^{2r+1},\quad \sum_{\ell=-\infty}^{\infty}\ell^{2r}.
\end{eqnarray}
We use a cutoff $N$ to perform the summation $\sum_{\ell}1$,
\begin{eqnarray}\label{appA2}
\sum_{\ell=-\infty}^{\infty}1&=&\lim\limits_{N\to \infty}\sum_{\ell=-N}^{N}1=\lim\limits_{N\to \infty}(1+2N)\nonumber\\
&=&1+\text{divergent term for $N\to \infty$}.
\end{eqnarray}
On the other hand,
\begin{eqnarray}\label{appA3}
\sum_{\ell=-\infty}^{\infty}\ell^{2r+1}=0,
\end{eqnarray}
and for $r\neq 0$
\begin{eqnarray}\label{appA4}
\sum_{\ell=-\infty}^{\infty}\ell^{2r}&=&\lim\limits_{N\to \infty} \sum_{\ell=-N}^{N}\ell^{2r}\nonumber\\
&=&0+\text{divergent term for $N\to \infty$}.
\end{eqnarray}
Neglecting the divergent terms in \eqref{appA2} and \eqref{appA4}, the summations in \eqref{appA1} are regularized as
\begin{eqnarray}\label{appA5}
\sum_{\ell=-\infty}^{\infty}1\to 1, \quad \sum_{\ell=-\infty}^{\infty}\ell^{2r+1}\to 0,\quad \sum_{\ell=-\infty}^{\infty}\ell^{2r}\to 0.\nonumber\\
\end{eqnarray}
As concerns \eqref{N17}, we first use
\begin{eqnarray}\label{appA6}
\ln(1+x)=\sum_{j=1}^{\infty}\frac{(-1)^{j+1}}{j}x^{j},
\end{eqnarray}
to rewrite $\ln(1+\alpha e^{\beta\ell\Omega})$ as
\begin{eqnarray}\label{appA7}
\mathcal{S}\equiv\sum_{\ell=-\infty}^{\infty}\ln(1+\alpha e^{\beta\ell\Omega})=\sum_{\ell=-\infty}^{\infty}\sum_{j=1}^{\infty}\frac{(-1)^{j+1}}{j}\alpha^{j}e^{\beta\ell\Omega j}. \nonumber\\
\end{eqnarray}
Then, plugging the Taylor expansion of $e^{\beta\ell\Omega j}$ into the r.h.s. of \eqref{appA7}, we arrive at
\begin{eqnarray}\label{appA8}
\mathcal{S}=\sum_{j=1}^{\infty}\frac{(-1)^{j+1}}{j}\alpha^{j}\sum_{r=0}^{\infty}\frac{\left(\beta\Omega j\right)^{r}}{r!}\sum_{\ell=-\infty}^{\infty}\ell^{r}.
\end{eqnarray}
In the regularization scheme \eqref{appA5}, we have
\begin{eqnarray}\label{appA9}
\sum_{\ell=-\infty}^{\infty}\ell^{r}=\sum_{\ell=-\infty}^{\infty}1+\sum_{\ell=-\infty}^{\infty}\ell^{2r}+\sum_{\ell=-\infty}^{\infty}\ell^{2r+1}\to 1.\nonumber\\
\end{eqnarray}
The first and second terms on the r.h.s. of \eqref{appA9} arise from $r=0$ and $r\neq 0$ contributions, respectively. Hence, the only nonvanishing contribution in \eqref{appA8} arises from $r=0$. This leads to
\begin{eqnarray}\label{appA10}
\mathcal{S}=\sum_{j=1}^{\infty}\frac{(-1)^{j+1}}{j}\alpha^{j}=\ln\left(1+\alpha\right),
\end{eqnarray}
as is claimed in \eqref{N17}.
\end{appendix}


\end{document}